\documentstyle[aps,preprint]{revtex}
\begin{document}
\draft
\title{Complete $\bbox{0\hbar\omega}$ calculations of Gamow-Teller
strengths\\
for nuclei in the iron region}

\author{D. J. Dean$^1$, P. B. Radha$^1$, K. Langanke$^1$, Y.
Alhassid$^2$,\\
S. E. Koonin$^1$, and W. E. Ormand$^1$\cite{Ormandadd}}
\address{$^1$ W. K. Kellogg Radiation Laboratory, 106-38, California
Institute of Technology\\ Pasadena, California 91125 USA\\
$^2$ Center for Theoretical Physics, Sloane Physics Laboratory\\
Yale University, New Haven, Connecticut 06511 USA}
\date{\today}
\maketitle

\widetext
\begin{abstract}
Gamow-Teller strengths for selected nuclei in the iron region
($A\sim56$) have been investigated via shell-model Monte Carlo
calculations with realistic interactions in the complete $fp$ basis.
Results for all cases show significant quenching relative to
single-particle estimates, in quantitative agreement with $(n,p)$
data. The $J=1,T=0$ residual interaction and the $f_{7/2}$--$f_{5/2}$
spin-orbit splitting are shown to play major roles in the quenching
mechanism. Calculated $B(E2, 2^+_1\rightarrow0^+_1)$ values are in
fair agreement with experiment using effective charges of $e_p=1.1~e$
and $e_n=0.1~e$.
\end{abstract}

\narrowtext

Gamow-Teller (GT) transitions from the ground states of medium and
heavy nuclei are important in both pre-supernova neutronization
\cite{SN} and double $\beta$-decay \cite{2beta}. These transitions
can be investigated experimentally through electron
capture/$\beta$-decay studies and through intermediate-energy $(p,n)$
and $(n,p)$ reactions. A long-standing puzzle has been that
experiments \cite{pn} systematically yield total GT strengths several
times smaller than single-particle estimates.

A number of explanations have been proposed for the observed
quenching of the GT strength, including a depletion of strength by
$\Delta$-excitations at about 300~MeV \cite{Delta} and a reduction of
the axial coupling constant to $g_A\cong1$ in nuclear matter
\cite{ga}. A less exotic cause would be the presence of
multi-particle, multi-hole configurations, and in light nuclei, where
complete $0\hbar\omega$ shell model calculations are possible, the GT
strength generally decreases as the model space is enlarged
\cite{Auerbach}. However, an exponentially increasing computational
difficulty limits similar studies in nuclei in mid-$fp$-shell to the
$2p$--$2h$ level, where calculations show a quenching only about half
of that observed experimentally.

While full $0\hbar\omega$ calculations apparently do not recover the
complete quenching of the GT-strength for $sd$-shell nuclei
\cite{ga}, there are phenomenological indications that a complete
treatment of the $fp$-shell is both necessary and sufficient to
describe GT quenching in the iron region \cite{Koonin}, and a first
calculation of ${}^{54}$Fe in such a basis with a realistic
interaction resulted in a quenching significantly larger than
truncated estimates and comparable to experiment \cite{Alhassid}. The
purpose of this Letter is to present calculations for several nuclei
in the iron region (${}^{54}$Cr, ${}^{55}$Mn, ${}^{54,56}$Fe, and
${}^{56,58}$Ni) aimed at exploring the universality of this quenching
and its dependence upon the effective interaction.

Our calculations are performed in the complete set of
$0f_{7/2,5/2}$--$1p_{3/2,1/2}$ configurations using the Monte Carlo
methods described in Refs.~\cite{Alhassid,Lang}. As in previous
truncated shell-model calculations of iron region nuclei
\cite{Auerbach}, we have used for the bulk of our studies the
Brown-Richter Hamiltonian \cite{Brown} fitted to lower $fp$-shell
nuclei, but we have also calculated ${}^{54}$Fe with the Kuo-Brown
interaction \cite{KB}. Calculations for ${}^{54,56}$Fe,
${}^{56,58}$Ni, and ${}^{54}$Cr were performed at $\beta=2~{\rm
MeV}^{-1}$ and 32 time slices (both of which were checked to be
sufficiently large to guarantee cooling to the ground state), while
our ${}^{55}$Mn results are at $\beta=1~{\rm MeV}^{-1}$ and 16 time
slices, evaluated using a weighted average of a ${}^{54}$Cr ensemble.
Each calculation involved some 3300 Monte Carlo samples at each of
six values of the coupling constant $g$ \cite{Alhassid} equally
spaced between $-1$ and 0; extrapolation to the physical case ($g=1$)
was done by the method described in Ref.~\cite{Alhassid}.

In Table~\ref{Tab1} we give results for selected static observables.
When the calculated binding energies are corrected for the Coulomb
energy using the semi-empirical formula
$E_c=0.717(Z^2/A^{1/3})(1-1.69/A^{2/3})$~MeV \cite{Nakayama2}, all
nuclei are overbound by about 1.5--2~MeV except ${}^{56}$Ni
($\sim3$~MeV) and ${}^{58}$Ni ($\sim0$~MeV). The
$B(E2,2_1^+\rightarrow0_1^+)$ values calculated with bare nucleon
charges are typically within a factor of 2 of the experimental values
if we assume that this transition saturates the total $0\hbar\omega$
strength. The agreement is improved if core polarization effects are
simulated by introducing effective charges that account for coupling
to $2^+$ configurations involving other major shells. We find that
adopting $e_p=1.1~e$ and $e_n=0.1~e$ for protons and neutrons,
respectively, reproduces the observed $B(E2)$ values for ${}^{54}$Fe,
${}^{56}$Ni, and ${}^{58}$Ni, while for the open-shell nuclei
${}^{54}$Cr and ${}^{56}$Fe slightly larger effective charges
($e_p\approx1.2~e$, $e_n\approx0.2~e$) are needed, indicating a
larger core polarization in the latter two than in the magic and
semi-magic nuclei of the $A\approx56$ region. These effective charges
are significantly smaller than those required in truncated shell
model calculations, indicating that the complete $0\hbar\omega$ model
space contains significant correlations absent in smaller model
spaces.

In Table~\ref{Tab2} we list results for the Gamow-Teller strengths
$B({\rm GT}_+)=\langle(\sigma\tau_+)^2\rangle$. In general, the ${\rm
GT}_+$ strength is quenched significantly relative to the
single-particle value, and is in good agreement with experiment in
all cases except ${}^{58}$Ni. Since the GT operators $\sigma\tau_+$
induce only $0\hbar\omega$ transitions, we do not expect a
significant renormalization of the operators. Note that the quenching
factors vary significantly from one nucleus to another and are not
well approximated by a common constant value, as is conventional in
astrophysical applications \cite{Mathews}. Compared to restricted
($2p$-$2h$) shell model approaches \cite{Auerbach}, our full
$fp$-shell calculation recovers about twice the quenching for both
${}^{54}$Fe and ${}^{56}$Ni.

The energy centroids of the ${\rm GT}_\pm$ strengths are also listed
in Table~\ref{Tab2} and can be compared with the experimental results
from $(n,p)$ and $(p,n)$ reactions \cite{pn}. Since we use
isospin-invariant Hamiltonians, analogue states are exactly
degenerate and the calculated ${\rm GT}_-$ centroid is the energy of
the GT resonance relative to the isobaric analogue state, $\Delta E=
E_{\rm GT_-}-E_{\rm IAS}$. Experimental values for the latter are
well parametrized by \cite{Nakayama} $\Delta E=(6.8-27.9\cdot
(N-Z)/A)$~MeV. Although $(N-Z)/A$ in our case is outside the range
studied by Nakayama {\it et al.} \cite{Nakayama}, we find that the MC
results in all of our cases agree with this parametrization to within
0.4~MeV, except for the double-magic nucleus ${}^{56}$Ni where
shell-closure effects apparently not considered in the empirical
parametrization might be important.

To compare the ${\rm GT}_{\pm}$ centroids to the data, we calculate
the excitation energy of the resonance $E_x$ in the daughter nucleus
either by using known analogue states or from the experimental mass
difference corrected by the Coulomb energy. Although the calculated
centroids for the $(n,p)$ reactions $({\rm GT}_+)$ are systematically
1.5 or 2~MeV too low, they generally track the position of the GT
resonance in the daughter nucleus well for all nuclei. In particular,
the calculation places the GT strength for the odd-$Z$ target
${}^{55}$Mn at a higher excitation energy $E_x$ than in the
neighboring nuclei, in accord with experiment. A similarly high
excitation energy is also observed for ${}^{51}$V and ${}^{59}$Co,
which aside from ${}^{55}$Mn, are the only odd-$Z$, mid-$fp$-shell
nuclei experimentally studied thus far, suggesting an odd-even
dependence of the excitation energy centroid. This might have an
important consequence for the measurement of the $B({\rm GT}_+)$
strength in $(n,p)$ reactions, which can determine the strength
reliably only up to daughter excitation energies of about 8~MeV.
Given the odd-even dependence, experiments with odd-$Z$ targets are
therefore likely to ``miss'' a relatively larger fraction of the
total strength with $E_x>8$~MeV, an effect that must be taken into
account if one wants to compare total strengths \cite{Koonin}. The
odd-even dependence should also be quite important in astrophysical
applications like pre-supernova calculations, but apparently has been
neglected to date.

It is important to determine the causes of the large GT quenching.
There is strong evidence that neutron-proton correlations in the
ground state are the source of the quenching, e.g.,
Ref.~\cite{Bertsch}. Engel {\it et al.} have suggested that the GT
quenching is particularly sensitive to the $T=0,J=1$ matrix element
\cite{Vogel}. To investigate this point, we show in Fig.~\ref{fig1}
$B({\rm GT}_+)$ for ${}^{54}$Fe in a calculation where all $T=0,J=1$
particle-particle matrix elements have been scaled by $0\leq
g_{pp}\leq 2$ (with $g_{pp}=1$ being the physical value). We show the
results for $g=0$ only, but we expect the extrapolated results at
$g=1$ to exhibit a similar behavior. There is a high sensitivity to
$g_{pp}$, and, upon comparing the result at $g_{pp}=0$ (all $T=0,J=1$
matrix elements vanishing) to that at $g_{pp}=1$, we see that this
component of the interaction causes about half of the total
quenching, in good agreement with the phenomenological prediction of
Ref.~\cite{Vogel}. In a further calculation, we switched off {\it
all} $T=0$ matrix elements and recovered about 90\% of the
single-particle estimate for $B({\rm GT}_+)$. Hino {\it et al.} have
argued that the increase of ground state correlations is accompanied
by an increase of $\langle J_p^2\rangle$ and $\langle J^2_n\rangle$
as proton-neutron states with $J\not=0$ will be admixed into the
ground state by $pn$-correlations \cite{Hino}. The increase of
$\langle J^2_{p,n}\rangle$ with increasing $g_{pp}$ shown in Fig.~1
supports this argument.

The GT quenching is also expected to be very sensitive to the
$f_{7/2}$-$f_{5/2}$ spin-orbit splitting. To investigate this effect,
we have modified the BR interaction by decreasing the $f_{5/2}$
single-particle energy arbitrarily by 2~MeV, thus reducing
$\epsilon_{5/2}-\epsilon_{7/2}$ from 6.49~MeV to 4.49~MeV (this
change defines the ``MBR'' interaction). $B({\rm GT}_+)$ values
calculated with this force are shown in Table~\ref{Tab3}, where we
also list effective spin-orbit splittings derived from the difference
in the energy centroids of the neutron spectral functions, $\langle
a^\dagger_{7/2}a_{7/2}\rangle$ and $\langle
a_{5/2}a^\dagger_{5/2}\rangle$. The effective spin-orbit splitting
generally tracks the single-particle energies, and is comparable to
the expected single-particle splitting of 7.16~MeV. It is clear from
these results that even modest changes in the single-particle
energies can have large effects on the quenching. Lowering the
$f_{7/2}-f_{5/2}$ energy splitting decreases the total binding energy
(which is overestimated by BR) significantly. However, it also
decreases the centroids of the GT strengths which were already too
low. Thus, the overbinding and the underestimation of the ${\rm
GT}_+$ excitation energy cannot be solved simultaneously by a shift
of the $f_{7/2}$-$f_{5/2}$ energy splitting.

We have also calculated ${}^{54}$Fe with the KB interaction. We find
a GT strength about a factor of two smaller than for the BR
interaction, indicating a strong, unanticipated sensitivity of the GT
quenching to the effective interaction. A similar dependence is also
observed in the $B(E2)$ values, for which the KB interaction predicts
a value 50\% larger than the BR interaction and in excess of the
data. As the single particle energies of the two interactions are
nearly equal, the observed differences must stem from differences in
the two-body interaction.

In summary, our complete $fp$-basis calculations of nuclei with
$A\sim56$ demonstrate that substantial quenching of the ${\rm GT}_+$
strength is a general phenomenon for all of the nuclei calculated and
that restricted shell model calculations miss a significant part of
this quenching. Our calculations with the BR interaction give a
quenching comparable to that observed in $(n,p)$ reactions; a
calculation of ${}^{54}$Fe with the KB interaction gives an even
greater quenching. Our results show that $np$ correlations are a
major contributor to this quenching, as is a small
$f_{7/2}$-$f_{5/2}$ spin-orbit splitting. The binding energies and
$B(E2, 2^+_1\rightarrow 0^+_1)$ values calculated with the BR
interaction are in reasonable agreement with experiment. In
particular, the latter require significantly smaller effective
charges than do previous, more restricted calculations.

Of course, all of our results depend upon the Hamiltonian used. The
BR interaction has been fitted to the lower $fp$-shell, and is
probably not optimal for the present studies. In fact, we find some
systematic shortcomings as the BR interaction tends to overbinding
while placing the GT centroids at somewhat too low an excitation
energy. Given these deficiencies and the strong sensitivity of the
${\rm GT}_+$ strength to the effective interaction, it is premature
to draw definite conclusions from the apparent good agreement between
the experimental and calculated quenching of the ${\rm GT}_+$
strength. A definite conclusion as to whether the quenching can be
totally recovered in complete $fp$-basis calculations must wait until
improved effective interactions are available. The new computational
techniques we have used in this work would allow such an interaction
to be determined in the complete $fp$ or even $fp\,g_{9/2}$ basis.

\acknowledgements

This work was supported in part by the National Science Foundation,
Grants No. PHY90-13248 and PHY91-15574, and by the Department of
Energy, Grant No. DE-FG-0291ER-40608. WEO acknowledges the Weingart
Foundation for financial support. We are grateful to P.~Vogel and
B.~A.~Brown for helpful discussions. Computational cycles were
provided by the Caltech Concurrent Supercomputing Consortium on the
Intel DELTA parallel supercomputer.

\begin{figure}
\caption{The Gamow-Teller strength $B({\rm GT}_+)$ and the
expectation values of $\langle J^2\rangle$ for protons (full circles)
and neutrons (open circles) as a function of the scaled strength of
the $T=0,J=1$ interaction, $g_{pp}$, defined in the text. The
calculation has been performed for ${}^{54}$Fe at $\beta=2~{\rm
MeV}^{-1}$ using the BR interaction and for the coupling constant
$g=0$.}
\label{fig1}
\end{figure}

\widetext
\begin{table}
\caption{Static observables calculated for selected nuclei with
$A\sim56$.}
\label{Tab1}
\begin{tabular}{c r@{}l r@{}l r@{}l r@{}l r@{}l r@{}l r@{}l}
\tableline
Nucleus &
\multicolumn{2}{c}{${}^{54}$Fe} &
\multicolumn{2}{c}{${}^{54}$Fe} &
\multicolumn{2}{c}{${}^{56}$Ni} &
\multicolumn{2}{c}{${}^{54}$Cr} &
\multicolumn{2}{c}{${}^{55}$Mn} &
\multicolumn{2}{c}{${}^{56}$Fe} &
\multicolumn{2}{c}{${}^{58}$Ni}\\ \tableline
Force &
\multicolumn{2}{c}{KB\tablenotemark[1]} &
\multicolumn{2}{c}{BR} &
\multicolumn{2}{c}{BR} &
\multicolumn{2}{c}{BR} &
\multicolumn{2}{c}{BR} &
\multicolumn{2}{c}{BR} &
\multicolumn{2}{c}{BR} \\ \tableline\tableline
$-\langle H\rangle$ (MC)\tablenotemark[2] &
\multicolumn{2}{c}{---}&
131.4 &\,$\pm 0.4$ &
145.2 &\,$\pm 0.6$ &
134.4 &\,$\pm 0.5$ &
141.8 &\,$\pm 1.0$ &
151.7 &\,$\pm 0.6$ &
164.4 &\,$\pm 0.7$ \\
(exp)\tablenotemark[2] &
\multicolumn{2}{c}{129.7} &
\multicolumn{2}{c}{129.7} &
\multicolumn{2}{c}{141.9} &
\multicolumn{2}{c}{132.0} &
\multicolumn{2}{c}{140.0} &
\multicolumn{2}{c}{150.2} &
\multicolumn{2}{c}{164.4} \\ \tableline\tableline
$\langle Q^2\rangle$ (${\rm fm}^4$)\tablenotemark[3]&
2560 &\,$\pm 83$ &
1482 &\,$\pm 84$ &
1572 &\,$\pm 13$ &
1408 &\,$\pm 64$ &
1447 &\,$\pm 55$ &
1819 &\,$\pm 91$ &
1674 &\,$\pm 18$ \\
$\langle Q_v^2\rangle$ (${\rm fm}^4$)\tablenotemark[3]&
368 &\,$\pm 31$ &
381 &\,$\pm 34$ &
380 &\,$\pm 3$ &
424 &\,$\pm 36$ &
384 &\,$\pm 24$ &
416 &\,$\pm 41$ &
520 &\,$\pm 7$ \\
$\langle Q_p^2\rangle$ (${\rm fm}^4$)\tablenotemark[3] &
718 &\,$\pm 29$ &
478 &\,$\pm 5$ &
487 &\,$\pm 3$ &
420 &\,$\pm 23$ &
&&
542 &\,$\pm 32$ &
528 &\,$\pm 6$ \\
$\langle Q_n^2\rangle$ (${\rm fm}^4$)\tablenotemark[3] &
749 &\,$\pm 26$ &
454 &\,$\pm 30$ &
487 &\,$\pm 3$ &
496 &\,$\pm 31$ &
&&
582 &\,$\pm 37$ &
569 &\,$\pm 7$ \\ \tableline
$B(E2)$~(MC)\tablenotemark[4] &
144 &\,$\pm 6$ &
96 &\,$\pm 1$ &
98 &\,$\pm 1$ &
84 &\,$\pm 5$ &
&&
108 &\,$\pm 6$ &
106 &\,$\pm 1$ \\
&
199 &\,$\pm 9$ &
129 &\,$\pm 1$ &
132 &\,$\pm 1$ &
114 &\,$\pm 6$ &
&&
148 &\,$\pm 9$ &
142 &\,$\pm 2$ \\
(exp)\tablenotemark[4] &
124 &\,$\pm 10$ &
124 &\,$\pm 10$ &
120 &\,$\pm 10$ &
174 &\,$\pm 8$ &
&&
196 &\,$\pm 8$ &
139 &\,$\pm 4$ \\ \tableline\tableline
$\langle M1^2\rangle (\mu^2_N)$\tablenotemark[5] &
11.9 &\,$\pm 0.4$ &
14.1 &\,$\pm 0.4$ &
17.7 &\,$\pm 0.2$ &
13.2 &\,$\pm 0.4$ &
21.6 &\,$\pm 0.4$ &
14.7 &\,$\pm 0.4$ &
17.5 &\,$\pm 0.4$ \\ \tableline
\end{tabular}
\tablenotetext[1]{The KB interaction has an undefined energy shift,
so that $\langle H\rangle$ is meaningless.}
\tablenotetext[2]{Binding energies (in MeV) relative to ${}^{40}$Ca.}
\tablenotetext[3]{$Q_p$ and $Q_n$ are the proton and neutron mass
quadrupole moments, calculated with oscillator length $b=1.97
(A/56)^{1/6}$~fm, while $Q=Q_p+Q_n$ and $Q_v=Q_p-Q_n$.}
\tablenotetext[4]{$B(E2,2^+_1\rightarrow 0^+_1)$ in units of $e^2{\rm
fm}^4$ calculated with the bare charges (upper row) and effective
charges $e_p=1.1e$ and $e_n=0.1e$ (lower row) and assuming that this
transition saturates $\langle Q^2\rangle$.}
\tablenotetext[5]{$M1$ is the magnetic moment operator assuming free
nucleon $g$-factors.}
\end{table}
\narrowtext

\widetext
\begin{table}
\caption{Gamow-Teller strengths for various nuclei with $A\sim56$.
Upper section: the Monte Carlo (MC) results for the total ${\rm
GT}_+$ strengths are compared with the single-particle values
and the experimental strengths extracted from $(n,p)$ data. The lower
two sections give the calculated energy
centroids of the GT strengths relative to the target ground state
$(\bar E_{\rm GT})$ and in the daughter nucleus $(E_x)$. The
experimental data are from \protect\cite{pn}. The quoted values for
$\bar E_{{\rm GT}_+}$ assume isospin conservation.}
\label{Tab2}
\begin{tabular}{c r@{}l r@{}l r@{}l r@{}l r@{}l r@{}l r@{}l }
\tableline
Nucleus &
\multicolumn{2}{c}{${}^{54}$Fe} &
\multicolumn{2}{c}{${}^{54}$Fe} &
\multicolumn{2}{c}{${}^{56}$Ni} &
\multicolumn{2}{c}{${}^{54}$Cr} &
\multicolumn{2}{c}{${}^{55}$Mn} &
\multicolumn{2}{c}{${}^{56}$Fe} &
\multicolumn{2}{c}{${}^{58}$Ni} \\ \tableline
Force &
\multicolumn{2}{c}{KB} &
\multicolumn{2}{c}{BR} &
\multicolumn{2}{c}{BR} &
\multicolumn{2}{c}{BR} &
\multicolumn{2}{c}{BR} &
\multicolumn{2}{c}{BR} &
\multicolumn{2}{c}{BR} \\ \tableline\tableline
$\langle {\rm GT}_+^2\rangle$ (sp) &
\multicolumn{2}{c}{10.3} &
\multicolumn{2}{c}{10.3} &
\multicolumn{2}{c}{13.7} &
\multicolumn{2}{c}{6.9} &
\multicolumn{2}{c}{8.6} &
\multicolumn{2}{c}{10.3} &
\multicolumn{2}{c}{13.7} \\
$\langle {\rm GT}_+^2\rangle$ (MC) &
2.2 &\,$\pm 0.3$ &
4.3 &\,$\pm 0.2$ &
7.4 &\,$\pm 0.3$ &
1.4 &\,$\pm 0.1$ &
2.2 &\,$\pm 0.2$ &
2.73 &\,$\pm 0.04$ &
5.6 &\,$\pm 0.3$ \\
$\langle {\rm GT}_+^2\rangle$ (exp) &
3.5 &\,$\pm 0.7$\tablenotemark[1] &
3.5 &\,$\pm 0.7$\tablenotemark[1] &
&&
&&
1.72 &\,$\pm 0.2$\tablenotemark[2] &
2.85 &\,$\pm 0.3$\tablenotemark[2] &
3.76 &\,$\pm 0.4$\tablenotemark[2] \\ \tableline\tableline
${\bar E}_{{\rm GT}_+}$ (MC) &
9.2 &\,$\pm 0.2$ &
9.7 &\,$\pm 0.2$ &
8.9 &\,$\pm 0.3$ &
&&
12.4 &\,$\pm 1.6$ &
11.0 &\,$\pm 0.2$ &
9.8 &\,$\pm 0.2$ \\
$E_x$ (MC)&
0.7 &\,$\pm 0.2$ &
1.2 &\,$\pm 0.2$ &
\multicolumn{2}{c}{2.7}&
&&
2.4 &\,$\pm 1.6$ &
-0.4 &\,$\pm 0.2$ &
1.2 &\,$\pm 0.2$ \\
$E_x$ (exp) &
\multicolumn{2}{c}{2.8} &
\multicolumn{2}{c}{2.8} &
&&
&&
4.1 &\,$\pm 0.5$ &
2.7 &\,$\pm 0.5$ &
3.5 &\,$\pm 0.5$ \\ \tableline
$\Delta E$ (MC)&
5.2 &\,$\pm 0.2$ &
6.1 &\,$\pm 0.2$ &
8.9 &\,$\pm 0.3$ &
&&
&&
4.4 &\,$\pm 0.1$ &
6.1 &\,$\pm 0.2$ \\
(sys)\tablenotemark[3] &
\multicolumn{2}{c}{5.8} &
\multicolumn{2}{c}{5.8} &
\multicolumn{2}{c}{6.8} &
\multicolumn{2}{c}{3.7} &
\multicolumn{2}{c}{4.3} &
\multicolumn{2}{c}{4.8} &
\multicolumn{2}{c}{5.9} \\
$E_x$ (MC) &
5.2 &\,$\pm 0.2$ &
6.1 &\,$\pm 0.2$ &
&&
&&
&&
7.8 &\,$\pm 0.1$ &
6.15 &\,$\pm 0.2$ \\
$E_x$ (exp) &
\multicolumn{2}{c}{8.2} &
\multicolumn{2}{c}{8.2} &
&&
&&
&&
&&
&\\
\tableline
\end{tabular}
\tablenotetext[1]{Summed to an excitation energy of 9 MeV.}
\tablenotetext[2]{Summed to an excitation energy of 8 MeV.}
\tablenotetext[3]{The systematics of Ref.~\protect\cite{Nakayama} for
$\bar E_{{\rm GT}_-}-E_{\rm IAS}$}
\end{table}
\narrowtext

\narrowtext
\begin{table}
\caption{Selected properties of ${}^{54}$Fe and ${}^{58}$Ni
calculated at $\beta=2$ with the Brown-Richter (BR) interaction of
Ref.~\protect\cite{Brown} and its modification (MBR), where the
$f_{5/2}$ single-particle energy has been lowered by 2~MeV.}
\label{Tab3}
\begin{tabular}{c r@{}l r@{}l r@{}l r@{}l}
Nucleus &
\multicolumn{2}{c}{${}^{54}$Fe} &
\multicolumn{2}{c}{${}^{54}$Fe} &
\multicolumn{2}{c}{${}^{58}$Ni} &
\multicolumn{2}{c}{${}^{58}$Ni} \\
Force &
\multicolumn{2}{c}{BR} &
\multicolumn{2}{c}{MBR} &
\multicolumn{2}{c}{BR} &
\multicolumn{2}{c}{MBR} \\ \tableline
$\langle {\rm GT}^2_+\rangle$ &
4.3 &\,$\pm0.2$ &
2.40 &\,$\pm0.03$ &
5.6 &\,$\pm0.3$ &
2.88 &\,$\pm0.04$ \\
$\bar E_{{\rm GT}_+}$&
9.7 &\,$\pm 0.2$ &
8.2 &\,$\pm 0.2$ &
9.8 &\,$\pm 0.2$ &
7.2 &\,$\pm 1.0$ \\
$-\langle H\rangle$&
131.4 &\,$\pm 0.4$ &
126.4 &\,$\pm 0.5$ &
164.4 &\,$\pm 0.7$ &
159.8 &\,$\pm 0.5$ \\ \tableline
$\bar E_{7/2}$\tablenotemark[1] &
11.77 &\,$\pm0.08$ &
12.36 &\,$\pm0.03$ &
11.73 &\,$\pm0.09$ &
10.40 &\,$\pm0.03$ \\
$\bar E_{5/2}$\tablenotemark[2] &
-3.45 &\,$\pm0.08$ &
-5.54 &\,$\pm0.08$ &
-3.36 &\,$\pm0.16$ &
-4.18 &\,$\pm0.09$ \\
$\Delta \bar E_{LS}$\tablenotemark[3] &
8.31 &\,$\pm0.11$ &
6.82 &\,$\pm0.08$ &
8.37 &\,$\pm0.18$ &
6.22 &\,$\pm0.09$ \\
\end{tabular}
\tablenotetext[1]{Energy centroid in MeV of the neutron $f_{7/2}$
strength, obtained from the response function $\langle
a^\dagger_{7/2} (\tau) a_{7/2} \rangle$.}
\tablenotetext[2]{Energy centroid of the neutron $f_{5/2}$ strength,
obtained from $\langle a_{5/2} (\tau) a^\dagger_{5/2} \rangle$.}
\tablenotetext[3]{Effective spin-orbit splitting, $\bar E_{7/2}+\bar
E_{5/2}$.}
\end{table}
\narrowtext

\end{document}